\documentclass{aa} 
\usepackage{graphicx}
\usepackage{aalongtable}
\begin{document}
   \title{Discovery of $\delta$ Scuti pulsation in the Herbig Ae star VV Ser}

   \subtitle{}

   \author{V. Ripepi\inst{1}\and
S. Bernabei\inst{2,3} \and
M. Marconi\inst{1}\and
A. Ruoppo\inst{1,4} \and
F. Palla\inst{5}\and 
M.J.P.F.G. Monteiro\inst{6}\and
J.P. Marques\inst{6}\and
P. Ferrara\inst{1}\and
S. Marinoni\inst{2}\and
L. Terranegra\inst{1}
          }

   \offprints{V. Ripepi}

   \institute{
INAF-Osservatorio Astronomico di Capodimonte,
Via Moiariello 16, 80131, Napoli, Italy \\
\email{ripepi@na.astro.it,marconi@na.astro.it,ruoppo@na.astro.it,pferrara@na.astro.it}
\and
INAF-Osservatorio Astronomico di Bologna, Via Ranzani 1,
40127 Bologna, Italy \\
\email{stefano.bernabei@bo.astro.it,silvia.marinoni@bo.astro.it}
\and
Departimento de Astrof\'{\i}sica, Universidad de La Laguna, Avda.
Astrofisico F. S\'anchez sn, 30071 La Laguna, Spain \\
\and
Dipartimento di Scienze Fisiche, Universit\`a Federico II, Complesso
Monte S. Angelo, 80126, Napoli, Italy \\
\and
INAF-Osservatorio Astrofisico di Arcetri, Largo E. Fermi, 5, I-50125
Firenze, Italy \\
\email{palla@arcetri.astro.it}
\and
DMA-Faculdade de Ci\^encias and Centro de Astrof\'{\i}sica da
Universidade do Porto, Rua das Estrelas, 4150-762 Porto, Portugal \\
\email{mjm@astro.up.pt}
             }

   \date{}


  \abstract
   {The study of pulsation in pre-main-sequence intermediate mass
stars represents an important tool to derive information on the
stellar parameters and structure, as well as to test the validity of
current theoretical models. The interest for this class of variable
stars has significantly increased during the last decade and about 30
members are presently known in the literature.}
   {A new observational study of the Herbig Ae star VV Ser has been
performed in order to detect and accurately measure pulsation
frequencies in the $\delta$ Scuti range, thus enlarging the sample of
known pulsators and contributing to the empirical definition of the
pre-main-sequence instability strip. As it belongs to the continuous
field of view of the asteroseismological satellite COROT, this study
also aims at characterizing the properties of VV Ser as a potential
``COROT additional program'' candidate.}
   {CCD time series photometry in the Johnson V filter has been
obtained during three consecutive years. The resulting light curves
have been subject to detailed frequency analysis and the derived
frequencies have been compared to model predictions.}
   {Seven pulsation frequencies have been measured on the basis of the
   best dataset obtained in 2004, ranging from $\sim$ 31 to $\sim$ 118
   $\mu Hz$, with an accuracy of the order of 0.5 $\mu Hz$. The
   comparison with an extensive set of asteroseismological models
   shows that all the observed periodicities can be reproduced if the
   stellar mass is close to $4 M_{\odot}$. Conversely, the measured
   frequencies can be associated to $p$ modes only if the effective
   temperature is significantly lower than that obtained from the
   spectral type conversion.}
{The present results seem to suggest that more accurate spectral
   type determination is necessary in order to discriminate the best
   fit model solution.  In any case, the stellar mass of VV
   Ser is close to the upper mass limit ($\sim 4 M_{\odot}$) for this
   class of pulsators.}


  \maketitle
%

\section{Introduction}

Pre-Main sequence (PMS) stars with masses larger than 1.5 M$_{\odot}$
are known as Herbig Ae/Be stars (\cite{herbig}). In general, they are
found within star forming regions and show variable emission lines
(especially H$\alpha$), as well as strong infrared excess caused by the
presence of circumstellar material (dust). 
In addition, Herbig Ae/Be stars are characterized by
photometric and spectroscopic
variability on time scales of minutes to years, mainly due
to photospheric activity and interaction with the circumstellar
environment (see e.g. \cite{gahm}, \cite{bohm1}). 

Considerable theoretical work has been done recently that
has advanced our understanding of PMS evolution (e.g. \cite{palla93};
\cite{dantona}, \cite{swenson}). Yet, there remain differences in
the models owing to alternative treatments of convection, opacities and
the zero-point of the calculated ages.
It is therefore desirable to find independent tools to constrain
PMS evolutionary tracks and, in turn, the internal structure of
intermediate-mass PMS stars.

Asteroseismology of Herbig Ae/Be stars can in principle allow us to
test PMS models by probing their interiors.
It is now well established that intermediate-mass PMS stars during
their contraction towards the MS cross the pulsation instability strip of
more evolved stars, giving origin to a variable class named
PMS $\delta$ Scuti stars (Kurtz \& Marang, 1995, Marconi \& Palla, 1998, 
\cite{catala}, \cite{ripepi05}).
The existence of pulsating Herbig Ae stars was originally proposed by
\cite{breger1} who discovered two candidates in the young open cluster
NGC 2264. More than 20 years later, the suggestion was confirmed by
\cite{kurtz} and \cite{donati} who observed $\delta$ Scuti-like
pulsations in the Herbig Ae stars HR~5999 and HD~104237, respectively.\par

The first theoretical investigation
of the PMS instability strip based on non-linear convective
hydrodynamical models was carried out by \cite{marconi} who calculated its
topology  for the first three radial modes. 
A subsequent theoretical work by \cite{suran}
made a comparative study of the seismology of a 1.8 $M_{\odot}$ PMS
and post-MS star. \cite{suran} found that the unstable frequency range
is roughly the same for PMS and post-MS stars, but that some
nonradial modes are very sensitive to the deep internal structure. 
In particular, it is possible to discriminate between the
PMS and post-MS phase using differences in the oscillation frequency
distribution in the low frequency range ($g$ modes). \par

Since the work by \cite{marconi}, many new PMS $\delta$ Scuti candidates have
been observed and the current number of known or suspected candidates
amounts to about 30 stars (see e.g. \cite{zwintz}, \cite{ripepi05}).
However, only a few stars have been studied in detail 
(e.g. \cite{v351}, \cite{bohm2}, \cite{ipper}), so that the overall
properties of this class of variables are still poorly determined.

In this context,  our group has started in the year 2000 a systematic
photometric monitoring program of Herbig Ae stars with spectral types
from A to F2-3 (see e.g. \cite{marconi2000}, \cite{h254}, \cite{v346},
\cite{v351}, \cite{ipper}). The aims were: 1) to identify
observationally the boundaries of the instability strip for PMS
$\delta$ Scuti pulsation; 2) to study in detail through multisite
campaigns selected objects showing multiperiodicity.
As part of this long-term project, we present here a new observational
study devoted to the Herbig Ae star VV Ser. In addition to the
motivation mentioned above, this star conveniently falls in the
continuous field of view of the asteroseismological satellite COROT
(corot.oamp.fr). Thus, VV Ser represents a potential ``additional program'' 
candidate
(\cite{weiss}).

In the next section, we discuss the properties of VV Ser; in Section 3
we describe the observations and data reduction; in Section 4 we deal
with the frequency analysis; in Section 5 we present a  comparison with
theoretical predictions. Finally, a brief summary closes the paper.

\begin{table}
\caption{Literature values of Spectral Type, Luminosity Class and 
rotational velocities for VV Ser}
\label{tab1a}
\begin{tabular}{ccc}
\hline
\noalign{\smallskip}
Sp. Type & v$\sin$i & Source \\
 & km/s &  \\
\hline
\noalign{\smallskip}
A2e  &   &  Herbig (1960) \\
A: &  & \cite{cohen} \\
B1e-B3e &  & \cite{FM84} \\
A2eV$\beta$ & & \cite{chavarria} \\
B9e$^{1}$ &  200$^{2}$ & \cite{hill1} \\
B5e  &   &  \cite{fernandez} \\
     & 85$^{3}$ &  \cite{grady} \\
A3IIe$\beta$ & & \cite{GC98} \\
A0Vevp$^{4}$  & 229 $\pm$ 9$^{5}$  & \cite{mora} \\
B+sh & 142 & \cite{vieira} \\
A2IIIe$^{6}$ & & \cite{acke} \\
B6e    & & \cite{hernandez}\\
\noalign{\smallskip}
\hline
\end{tabular}\\
1=photometrically determined \par
2=no reference is indicated for this value\par
3=the Authors quote Hillenbrand et al. (1995, private communication)
as source for this value\par
4=the Authors quote an error of five classes\par
5=v$\sin$i derived using just one line (MgII 4481 \AA~line)\par
6=the Authors quote van den Ancker (private communication) as
source for this value
\end{table}

\begin{table}
\caption{Stellar parameters for VV Ser found in the literature}
\label{tab1b}
\begin{tabular}{ccccccc}
\hline
\noalign{\smallskip}
V    & D  & $A_V$ &  $\log(T_\mathrm{eff})$ & $\log(L)$   & M            & source\\
mag  & pc & mag   &       K           & $L_{\odot}$ &  M$_{\odot}$ & \\
\hline
\noalign{\smallskip}
12.666 & 245 & 6.1 & 3.95 & $>$ 1.62 &       & 1  \\
11.87  & 440 & 3.0 & 4.03 & 1.8      &  3.3  & 2  \\
11.92  & 440 & 3.4 & 4.14 & 2.23     &  3.8  & 3  \\
11.63  & 296 & 3.4 & 3.95 & 1.51     &  2.1  & 4  \\
11.58  & 440 &     & 4.03 & 2.03     &       & 5  \\
       & 330 & 2.7 & 3.95 & 1.27     &       & 6  \\        
\noalign{\smallskip}
\hline
\end{tabular}
1=\cite{chavarria};2=\cite{hill1};\\ 
3=\cite{hernandez}; 
4=Rostopchina et al. (2001);\\
5=Testi et al. (1998); 6=\cite{acke}        
\end{table}

\section{Stellar parameters of VV Ser}

The variable star VV Ser was identified as a young Ae star in the
seminal work by Herbig (1960) on emission line stars associated with reflection nebulae.
The star is located in the Serpens molecular cloud
(Chavarria et al. 1988) and has been widely studied in literature,
although its properties and 
position in the HR diagram are still rather uncertain . In the following, we
present an overview of the current observational status of VV Ser.

\begin{itemize}

\item
{\bf Spectral type and luminosity class}: We report in
Table~\ref{tab1a} (first column) all the independent measurements of
spectral type (ST) found in literature. The
uncertainty is large with ST varying between B1e-B3e and A3e. The
determination of the earliest type by \cite{FM84} is based on the HeI
5876 \AA~ line which was later found to be associated with the hot
regions of accretion disks (see \cite{rostopchina} and references
therein) and therefore should not be considered reliable.  However, a
B5--B9 type has also been claimed by other authors (see Table~\ref{tab1a}).
On the other hand, Table~\ref{tab1a} lists six
studies where VV Ser is classified as an Ae star, with typical value
A2. Interestingly, the only paper exclusively dedicated to VV Ser is
that by \cite{chavarria} who assigns an A2e class, based on high
resolution spectroscopy and Str\"omgren photometry. In the following,
we will rely on these results, allowing an
uncertainty of 2 subclasses to take into account the spread of
spectral types found in the literature.  \par Concerning the
luminosity class of VV Ser, there are three determinations in the
literature: class II (Gray \& Corbally 1998), III (Acke \& van den
Ancker 2004), and V (Mora et al. 2001). Again, there is no agreement
between different authors.

\item
{\bf Distance and position in the HR Diagram}: The distance estimate
of VV Ser is related to that of the Serpens Cloud. 
On the basis of photometric and 
spectroscopic observations of 5 stars
in the cloud (excluding VV Ser), \cite{chavarria} and \cite{delara}
calculated a distance of D=311$\pm$38
pc, in fair agreement with previous determinations of about 440 pc by
\cite{racine} and \cite{strom}. In the following, we will adopt a 
distance of 300-400 pc. \par

As for the position in the HR diagram, Table~\ref{tab1b} lists several 
estimates of $\log(T_\mathrm{eff})$ and $\log(L/L_{\odot})$ found 
in literature. The table highlights the existing uncertainties, mainly because of
different assumptions on the ST, on the visual absorption and on the 
apparent visual magnitude of VV Ser which varies by tenth of magnitude
on time scales of weeks.

\item
{\bf Rotational velocity}: Another source of uncertainty in the
analysis of VV Ser is represented by its rotational velocity.
Table~\ref{tab1} reports four independent measurements of $v\sin i$ found
in the literature. The values reported by \cite{hill1} and \cite{mora}
are in fair agreement, whereas those given by \cite{grady} and
\cite{vieira} are significantly different.  More in detail, we were
not able to find the source of the $v\sin i$ reported by \cite{hill1}
who quoted the studies of \cite{finke1} and \cite{davis} that however
do not give a value for VV Ser.  Similarly, the low $v\sin i$ reported
by \cite{GC98} cannot be verified because it is quoted as private
communication.  Finally, the $v\sin i$ measured by \cite{mora} is
somewhat uncertain because it relies on just one line, the MgII 4481
\AA~line. \par Thus, the rotational velocity of VV Ser is still
uncertain, even if there are indications that it could be larger than
100 km/s. This possibility will be taken into account 
in the interpretation of the data (see Section~5).



\end{itemize}

In conclusion, the empirical estimate of the stellar parameters of VV
Ser is still plagued by large errors. However, as we will discuss in
Sect. 5, the pulsational analysis of the detected frequencies can
help to reduce the uncertainty on luminosity, effective
temperature and mass.

\section{Observations and data reduction}

The observations of VV Ser have been carried out in several runs
during 2002, 2003 and 2004 (a detailed observational log is 
given in electronic form in Table~\ref{tab1}), 
using the Loiano 1.5m telescope equipped with the
BFOSC instrument (see http://www.bo.astro.it/loiano/index.htm for a
detailed description of the telescope and of the instruments). The CCD
is a EEV 1340x1300 pixels with individual size of 0.58 arcsec, for a
total field of view of 13$^{'} \times 13^{'}$. In total, about 55
hours of observations have been gathered in the Johnson V filter.

The comparison star used was USNO-A2.0 0900-12930815 ($18^h$ $28^m$
$33.535^s$ $+00^{\circ}$ $02^{'}$ $25.71^{''} J2000$) which is about
7.3$^m$ away from VV Ser, allowing to observe both stars in a single
BFOSC frame.  We note that this star has already been used as
comparison for VV Ser by \cite{fernandez1} who found it stable at a
level of 0.02 mag and with an apparent magnitude and color of
$V=12.28$ mag $(B-V)=0.89$ mag, respectively.

   \begin{figure}
   \centering
   \includegraphics[height=8cm]{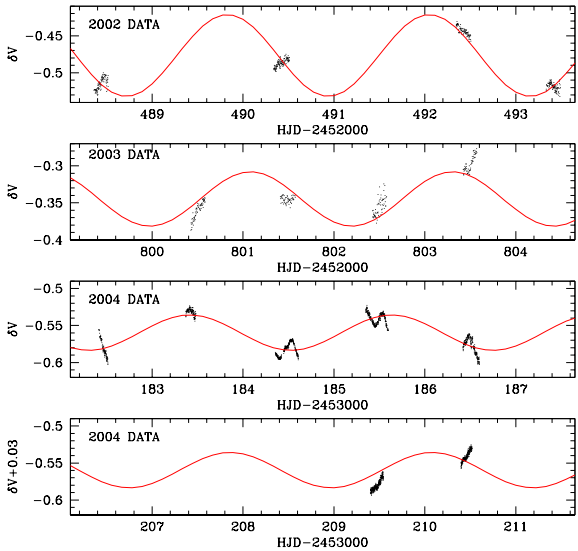}
      \caption{From top to bottom, the first two panels show the
 2002 and 2003 dataset; the third and fourth panel show
the 2004 dataset. Note that in the latter we have added 0.03 mag
to the original differential photometry (see Sect. 3).
The solid line shows the fit with a  $\sin$ curve with fixed period of 2.2 d
and free amplitude and phase (for each dataset, see Sect.3).}
      \label{fig1}
   \end{figure}
%

   \begin{figure*}
   \centering
\hbox{
   \includegraphics[height=9.2cm]{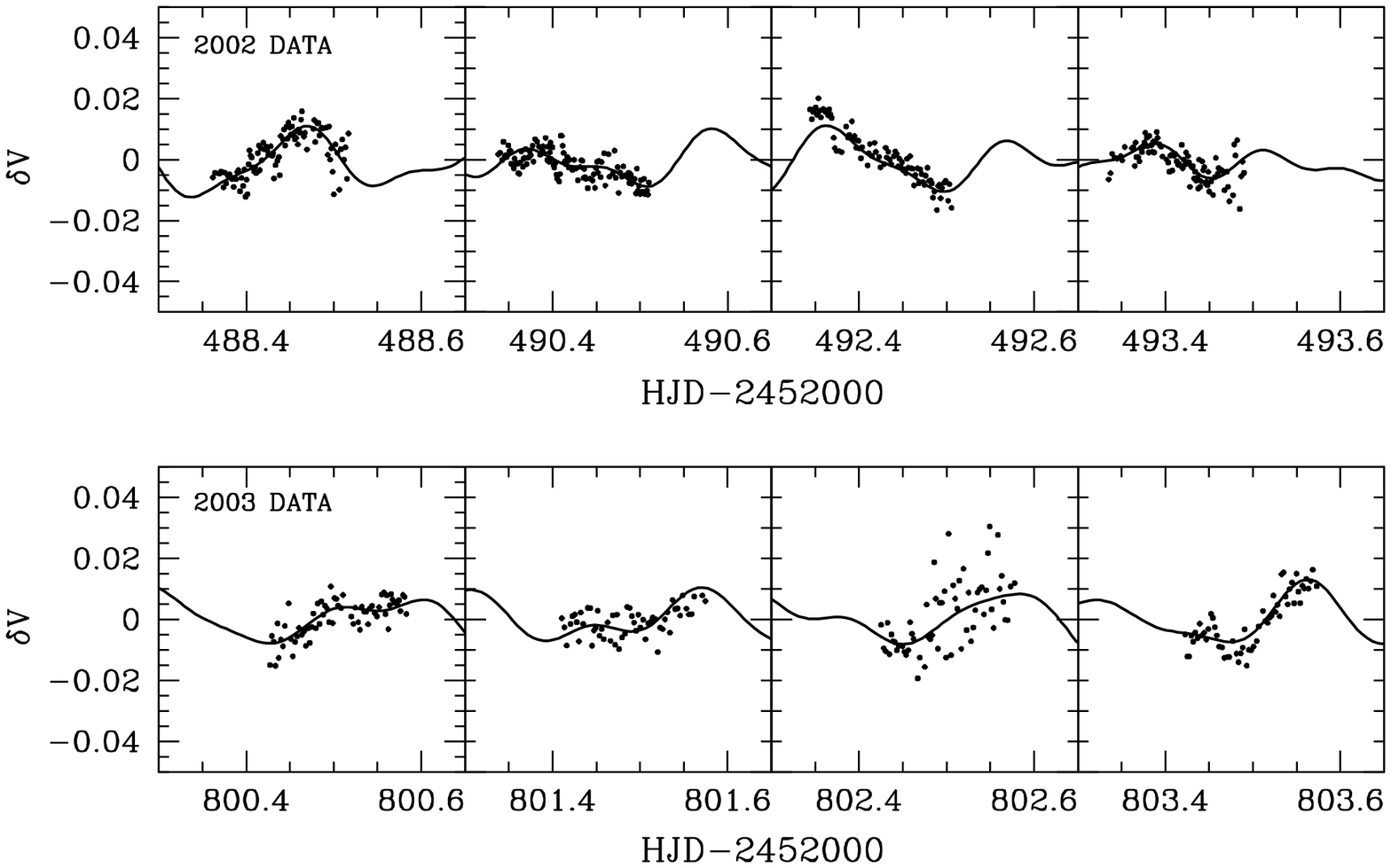}
   \includegraphics[height=9.2cm]{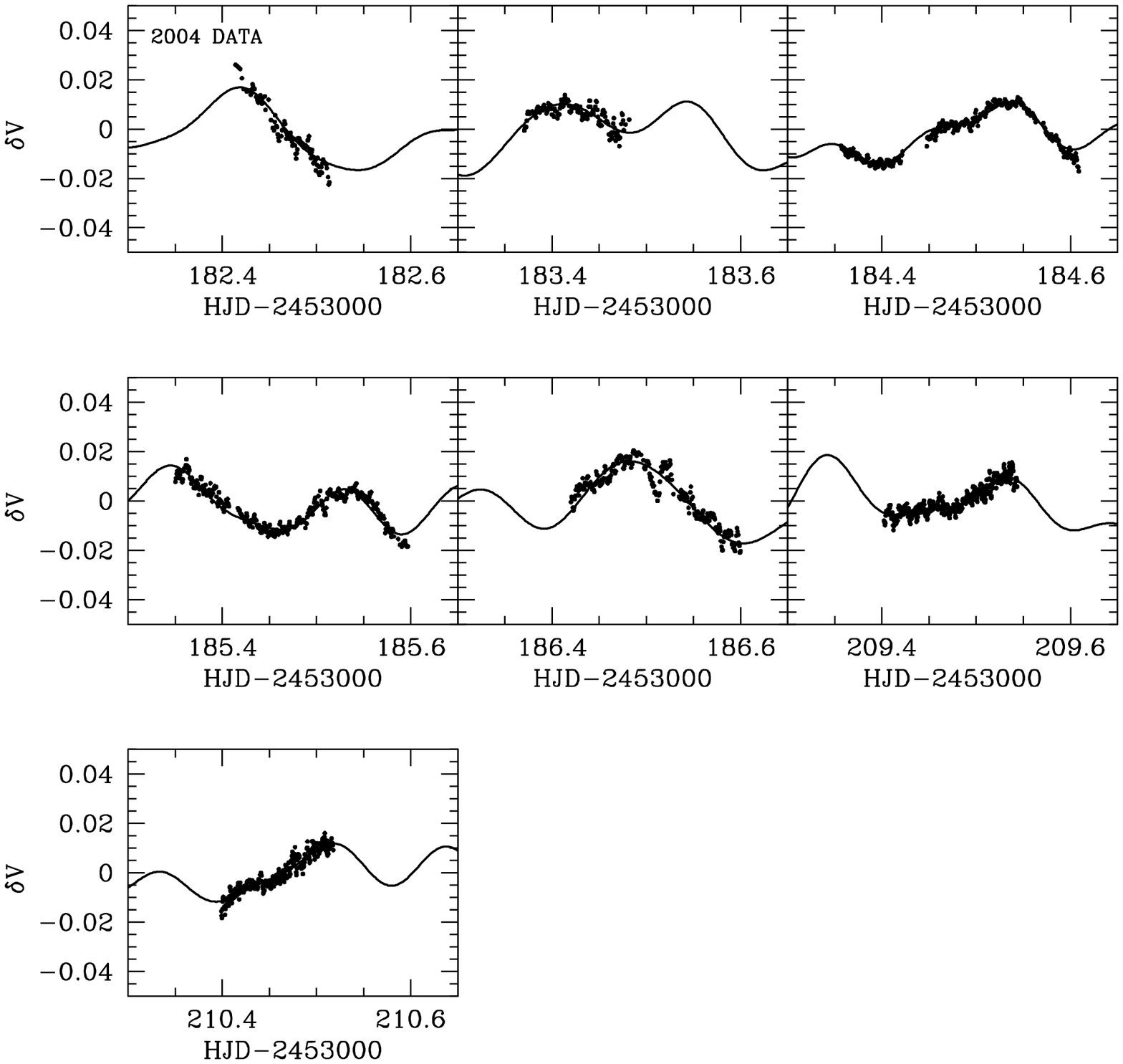}
}
      \caption{{\bf Left panels:} light curves for the 2002 (top)
and 2003 (bottom) datasets after the detrending. Note that
$\delta V=V_{VAR}-V_{COMP}$. {\bf Right panels:} as before, but for the 2004 
dataset. In all the panels, the solid line displays the fit to the
data with all the significant frequencies found for each dataset,
as listed in Table~\ref{tab3}.}
      \label{fig2}
   \end{figure*}
%

All the data have been reduced following the usual procedures
(de-biasing, flat-fielding) by using standard IRAF routines. The
aperture photometry has been carried out using routines written in the
MIDAS environment. The typical precision of the data was of about 4,
6, and 3 mmag for the 2002, 2003 and 2004 dataset, respectively. On
this basis (see also Fig.~\ref{fig1} and ~\ref{fig2}) it is evident
that the 2002 and especially the 2004 datasets are the best ones, whereas
the 2003 run is of poor quality. Note that, because of  the  high sampling, the
2004 data have been smoothed by using a 3-point boxcar filter.

The observed light curves for the different datasets are reported in
Fig.~\ref{fig1}. It is evident that VV Ser displays
mid-term variations on time scale of days, in addition to the pulsation
on time scale of hours. In order to eliminate these night-to-night
variations and to prepare the data for Fourier analysis, we decided to
detrend the data to a common average zero value. Then, we used
the period04 package (\cite{lenz}) to remove the frequencies 0.60 c/d
from the 2002 dataset, 0.46 c/d+0.19 c/d from the 2003 dataset, and
0.45 c/d from the 2004 dataset, with the exception of the two nights
with HJD=53209 and 53210 for which a simple average was subtracted.
As a result, we show in Fig.~\ref{fig2} the light curves for the three
datasets after the detrending procedure.\\ 
Before proceeding with the
frequency analysis of the obtained time series, we would like to
discuss more in detail the mid-term light variations with period of
the order of 2 d which have been removed from the data as described
before.  To investigate further this point, we fitted a sinusoid with
the period P=2.2 d (frequency 0.45 c/d) found using the 2004 data
(HJD=53182-53186) to the three datasets, leaving the amplitude and
phases as free parameters (to this aim we added 0.03 mag to the photometry 
obtained during the nights with HJD=53209 and 53210).  
The result is shown in Fig.~\ref{fig1}
by a solid line: the fit is fairly good also for
the 2002 dataset, whereas it is worse for the 2003 dataset. 
In conclusion, a periodicity of the
order of 2 d seems to be present in VV Ser. It is worth to note that
analysing six year data \cite{shevchenko} found both a long-term
periodicity (around 1000 d) and a short-term periodicity (which they
quote as uncertain) of 3.9 d. Even if there is a discrepancy between
\cite{shevchenko}'s results and ours, the short-term periodicity could
be real. The physical origin of this kind of periodicity is not easily
understood and in any case is not the subject of this paper. However, such
short-term variability in Herbig Ae stars is generally attributed to
rotational modulation due to magnetic activity (see
e.g. \cite{catala99}).

\section{Frequency analysis}

\begin{figure}
   \centering
   \includegraphics[height=9cm]{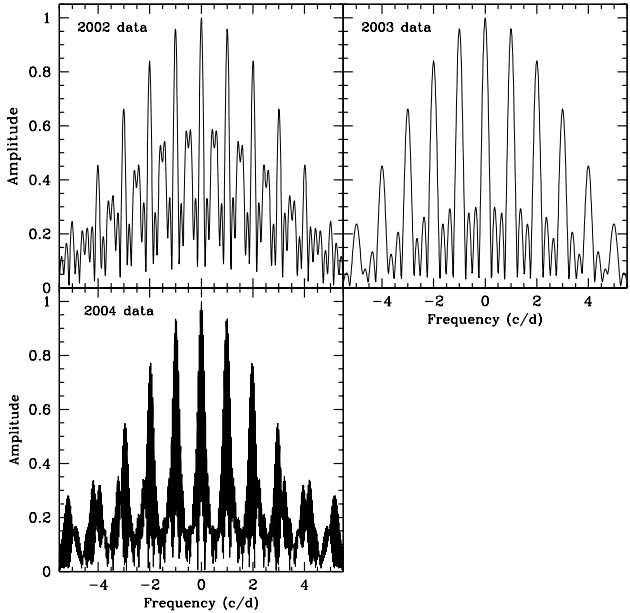}
      \caption{Spectral window in amplitude for the three data sets
analysed in this paper (see labels in the figure).}
      \label{fig3}
   \end{figure}

The detrended time series shown in Fig.~\ref{fig2} have been analysed
by using again the period04 package (\cite{lenz}), based on the
Fourier transform method. For a better interpretation of the results,
we have first calculated the spectral window (SW) for each
dataset. The result is shown in Fig.~\ref{fig3} where from top to
bottom we report the SW for the three time series identified in the
previous section (see labels in the figure). The SW was used as a
diagnostic to distinguish between real and spurious frequencies.

%

\begin{table}
\caption{Frequencies, amplitudes and confidence levels for the three
datasets analysed in this paper. The error on the 2002, 2003, and 2004
data are $\sim$0.14 c/d, $\sim$0.22 c/d, $\sim$ 0.04 c/d, respectively.
Note that for the 2002 and 2003 datasets the frequencies are labeled
in correspondence to the 2004 dataset (see text).}
\label{tab3}
\begin{tabular}{cccc}
\hline
\noalign{\smallskip}
  & Frequency   &   Amplitude  & Confidence  \\
  & (c/d)       &    (mmag)    & (\%)     \\
\noalign{\smallskip}
\hline
\multicolumn{4}{c}{\bf 2002 DATA} \\
\noalign{\smallskip}
 $f_1$ & 5.11 & 6.3   & 99.9 \\
 $f_4$ &	3.84 & 4.6   & 99.9 \\
 $f_5$ &	9.55 & 2.9   & 99.9 \\
\noalign{\smallskip}
 \multicolumn{4}{c}{\bf 2003 DATA} \\
\noalign{\smallskip}
 $f_4$&	3.99&  7.6 & 99.9 \\
 $f_5$&	8.50&  2.3 & 99.0 \\
 $f_1$&	6.11&  3.1 & 90.0 \\
\noalign{\smallskip}
 \multicolumn{4}{c}{\bf 2004 DATA} \\
\noalign{\smallskip}
$f_1$ &6.12	& 6.4  & 99.9 \\
$f_2$ &2.69	& 7.8  & 99.9 \\
$f_3$ &4.47	& 3.7  & 99.9 \\
$f_4$ &3.90	& 4.8  & 99.9 \\
$f_5$ &9.96	& 2.7  & 99.9 \\
$f_6$ &7.56	& 2.1  & 99.9 \\
$f_7$ &10.24	& 1.4  & 90.0 \\
\noalign{\smallskip}
\hline
\end{tabular}
\end{table}

\begin{figure*}
   \centering 
\hbox{
\includegraphics[width=9.2cm]{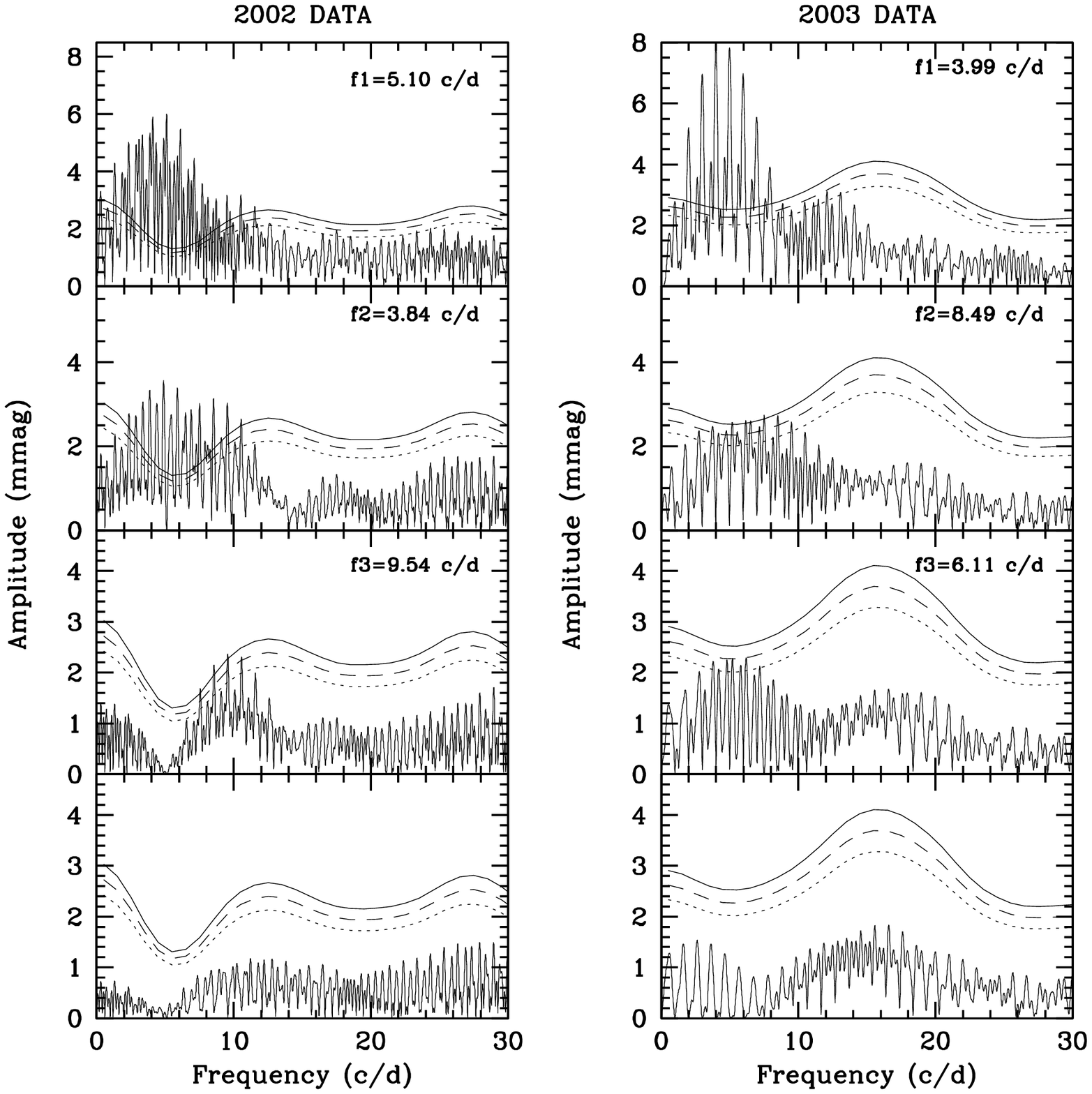}
   \includegraphics[width=9.2cm]{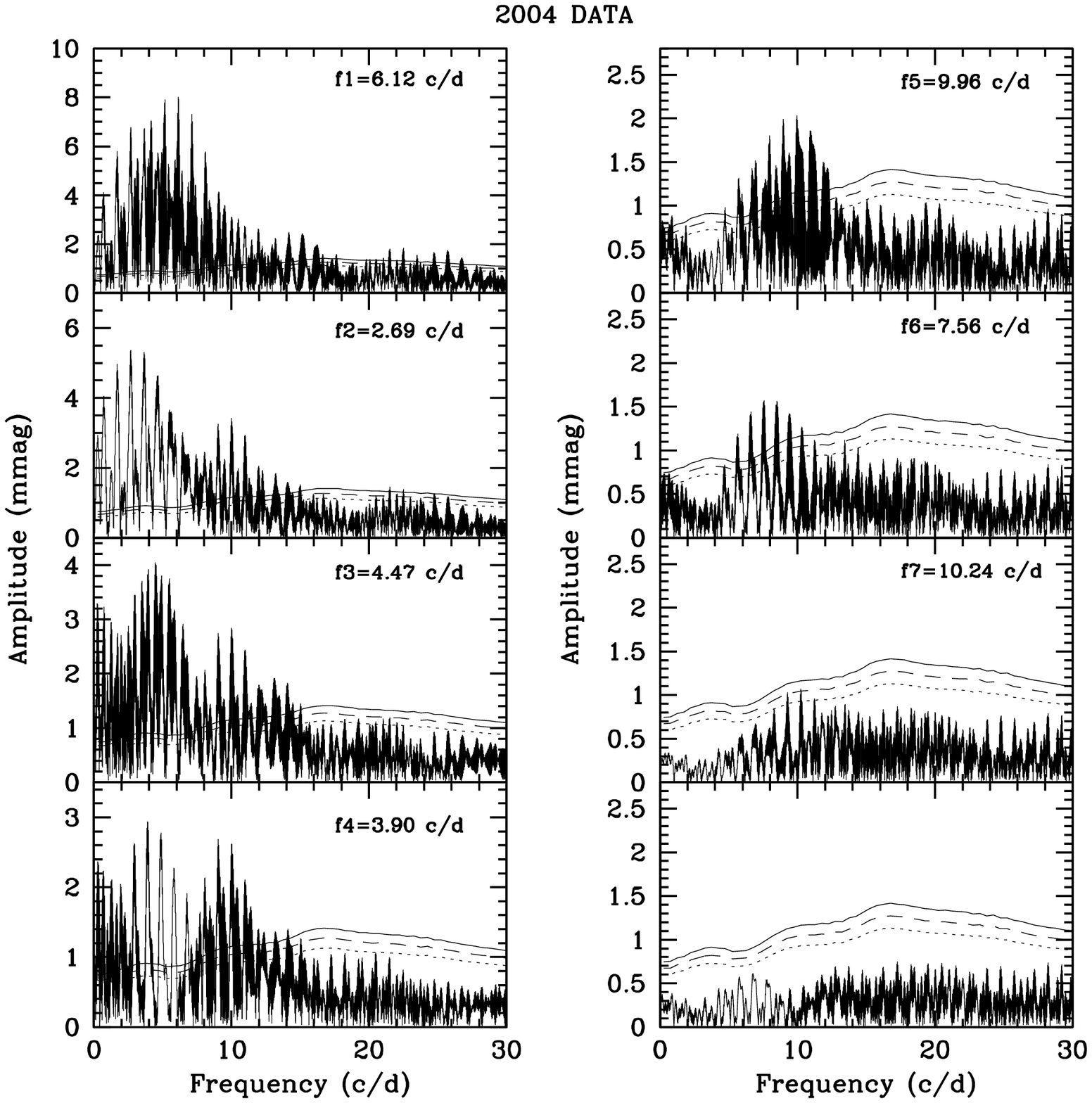}
}
      \caption{{\bf Left panels:} Frequency analysis for the 2002 and 2003 datasets.  The
solid, dashed and dotted lines show the 99.9\%, 99\% and
90\% significant levels. In each panel, one peak (i.e. the labelled
frequency) is selected and removed from the time series and a new
spectrum is obtained.  The last panel displays the periodogram after the
prewithening with all the significant frequencies. {\bf Right panels:} As before, 
but for the 2004 dataset. }
      \label{fig4}
   \end{figure*}

Each data set described in the previous Section has been analysed
separately. Figure~\ref{fig4} (left panels) shows the Fourier
transform for the 2002 and 2003 datasets, whereas Fig.~\ref{fig4}
(right panels) displays the 2004 one. Here, in each panel the peak with
largest amplitude is selected and then removed, obtaining a new
spectrum shown in the next panel.  The last panel shows the
periodogram after the prewithening with all the significant
frequencies. The solid, dashed and dotted lines represent the 99.9\%, 99\%
and 90\% confidence levels calculated following the widely used recipe
by \cite{breger93} and \cite{kus97}. The error on the measured
frequencies (apart from the $\pm$ 1 c/d alias) can be roughly
estimated from the FWHM of the main lobe in the spectral window (see
\cite{alvarez} and references therein). As a result, we found $\Delta
f$ $\sim$ 0.14 c/d, 0.22 c/d and 0.04 c/d for the 2002, 2003
and 2004 datasets. For the last dataset, due to the 
complex structure of the main lobe of the spectral window, in order to be 
conservative, we doubled the nominal error. It is important to emphasize 
that all the frequencies reported in the following are affected by the 
indetermination due to the 1 c/d alias.\par 

All the significant frequencies found for the three datasets are
summarized in Table~\ref{tab3}. A fit to each dataset with the quoted
frequencies is shown in Fig.~\ref{fig2} by a solid line. \\ In
order to discuss in detail the results of the frequency analysis, we will
take as reference the frequencies obtained with the best dataset,
that of 2004.Then, we find:

\begin{itemize}

\item
$f_1=6.12$ c/d: it is found in all datasets within the quoted
errors. The first frequency in 2002 is clearly the -1 c/d alias (this
periodicity has also the same amplitude as in 2004), whereas in 2003
it has a significantly lower amplitude, owing to the poor quality
of the dataset and/or to a possible change in amplitude (common
occurrence among $\delta$ Scuti stars)

\item
$f_2$, $f_3$, $f_6$, $f_7$: they  have been found only in the 2004 dataset. 
We note that $f_7$ would be exactly twice $f1$ if we took the -1 c/d alias, 
therefore it could be a non-independent frequency. 
Moreover it is only partially significant (90\%).

\item
$f_4=3.90$ c/d : within the errors, it can be considered the same frequency
as in the 2002 ($f_4=$3.84 c/d) and 2003 ($f_4=$3.99 c/d) datasets.

\item

$f_5=9.96$ c/d: it could  correspond to the frequencies at 9.55 c/d
and 8.50 c/d (-1 c/d alias) in the 2002 and 2003 datasets.
Although these three frequencies are not equivalent within
the errors, we will assume
that they represent the same frequency because of the similar amplitudes.

\end{itemize}

In the following, we will use
the set of 6 significant frequencies extracted from the 2004 dataset for
comparison with theoretical models.


%

%

%
\section{Comparison with theory}

We now attempt to use the frequency data reported
above to carry out a preliminary comparison of the available observables
with theoretical models.  To do so, we consider the observed location
of the star in the theoretical HR diagram and how a tentative mode
identification of the observed frequencies can constrain the stellar
mass.  Because the outcome depends strongly on
the assumptions, we will also evaluate the effects of the
uncertainties on the results.

\subsection{Location in the HR diagram}

\begin{figure}
   \centering
    \includegraphics[width=9cm]{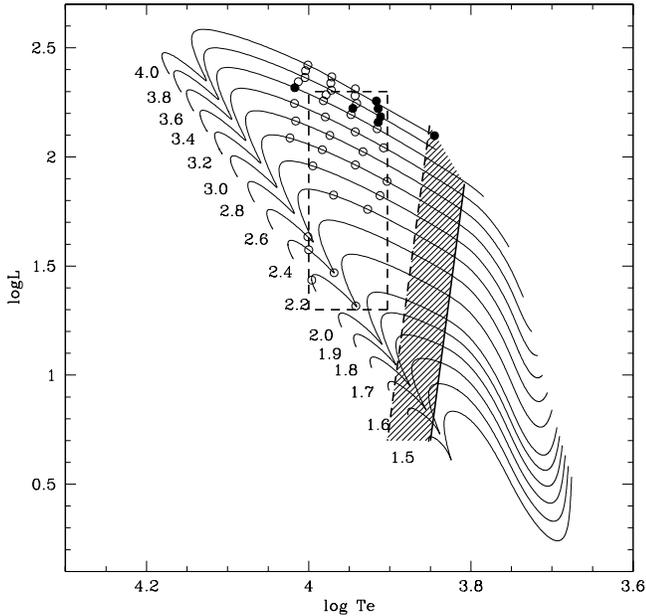}
      \caption{PMS evolutionary tracks for the labelled masses and
      solar chemical composition ($Z=0.02$, $Y=0.28$) computed with
      the CESAM code.  The dashed box represents the estimated uncertainty
      on the position of VV Ser. It is also shown the location of the
      theoretical instability strip for the first three radial modes
      predicted by \cite{marconi} on the basis of nonlinear convective
      pulsation models. The empty circles represent the 32 models run
      for the pulsational analysis. The filled symbols are for the
      best fit models discussed in Section 5.}
      \label{fig6}
   \end{figure}

 In order to compare the observations with the theoretical
predictions, it is necessary to estimate the position of VV~Ser in the
HR diagram. As discussed in Section~2, spectral type and distance to
VV Ser are quite uncertain.  First, we have estimated the $T_\mathrm{eff}$
value from the adopted spectral type A2$\pm$2 subclass by using the
\cite{kaler} conversion tables, noting that for these relatively
early-types these values are in agreement with the more recent
estimates of, e.g., \cite{kenyon} (see their Table A5) .  As a
result, we obtain $T_\mathrm{eff}$=9000$\pm$1000~K. 
Concerning the luminosity, we decided to investigate all
the values in the range $1.3<\log{L/L_{\odot}}<2.3$, covering
the empirical estimates in the literature.  The studied region is represented by
the large box in the HR diagram shown in Fig.~\ref{fig6} that
includes a set of PMS evolutionary tracks computed using the CESAM
code (\cite{morel}, \cite{marques}) for stellar masses from 1.5 to 4.5
$M_{\odot}$, and the theoretical instability strip computed for the
first three radial modes by \cite{marconi}, on the basis of nonlinear
convective pulsation models.  We note that for the estimated 
effective temperatures of VV~Ser we do not expect pulsation in the first three
radial modes since the predicted instability strip is cooler at
each luminosity level.  We might instead expect pulsation in higher
radial overtones and/or in nonradial modes. In fact the instability strip 
is predicted to move toward higher effective temperatures as 
the radial overtone increases (see Grigahc\`ene et al. 2006).

%

\subsection{Oscillation frequencies}

In order to evaluate the sensitivity of the predicted
periodicities to the input stellar parameters, we computed a fine grid
of structure models along the CESAM evolutionary tracks covering a
mass range from 2.2 to 4.0 $M_{\odot}$ and effective temperatures
from $\sim$ 8000 K to $\sim$ 10,000 K. The physical
properties of the
selected PMS models are reported in Table~\ref{tab4} 
(available in electronic form).

\begin{figure}
   \centering
   \includegraphics[width=9cm]{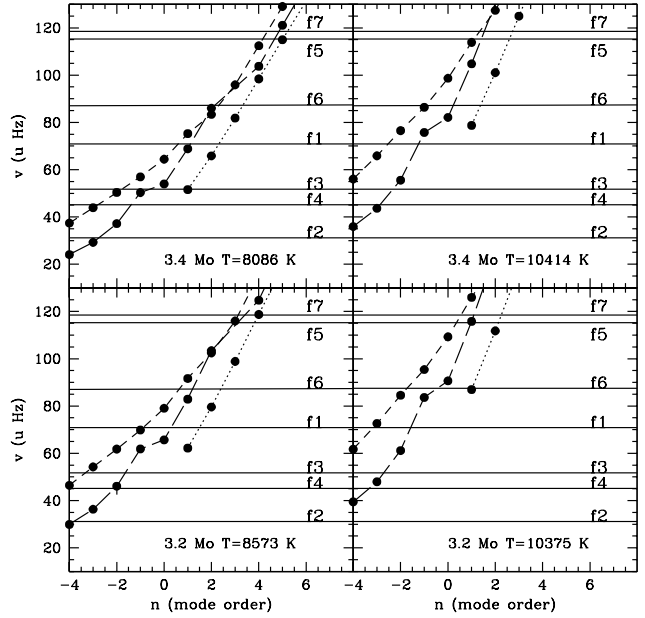}
      \caption{Comparison between predicted (filled symbols) and
      observed (horizontal solid lines) periodicities for models at
      3.2~$M_{\odot}$ (bottom panels), and 3.4~$M_{\odot}$ (upper
      panels).  Symbols are connected by a dotted line in the case of
      $l$=0, by a long-dashed line for $l$=1 and by a dashed line for
      $l$=2.  The two panels refer to the labelled effective
      temperatures.  When a predicted frequency reproduces an observed
      one the corresponding filled symbol (along the given l slanting
      line) is located at the intersection with the horizontal line
      representing the observed value.}
      \label{fig7}
   \end{figure}

\begin{figure}
   \centering
   \includegraphics[width=9cm]{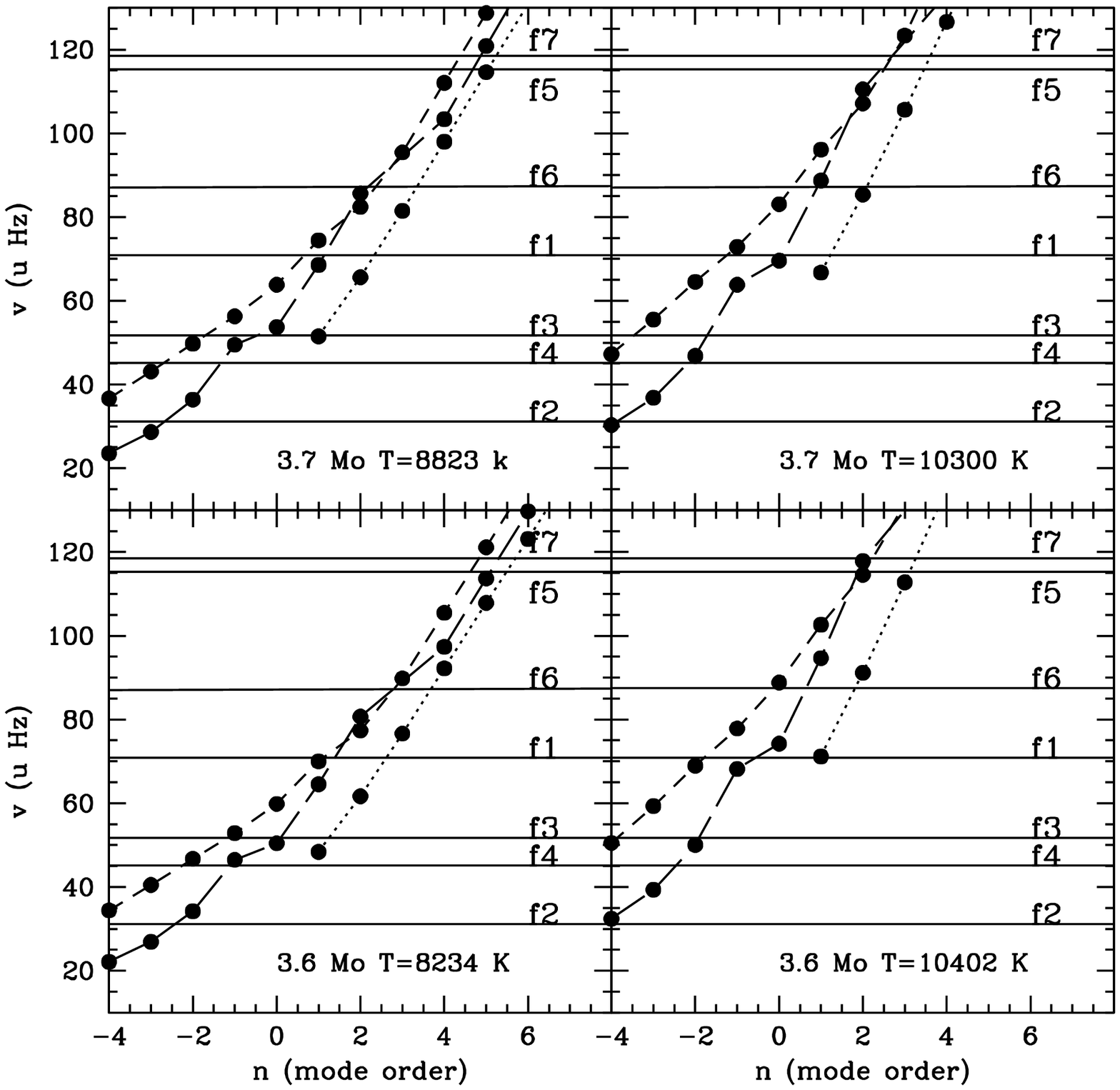}
      \caption{The same as for Fig.~\ref{fig7} but for 3.6~$M_{\odot}$ 
       (bottom panels)
      and 3.7~$M_{\odot}$ (upper panels).}
      \label{fig8}
   \end{figure}

\begin{figure}
   \centering
   \includegraphics[width=9cm]{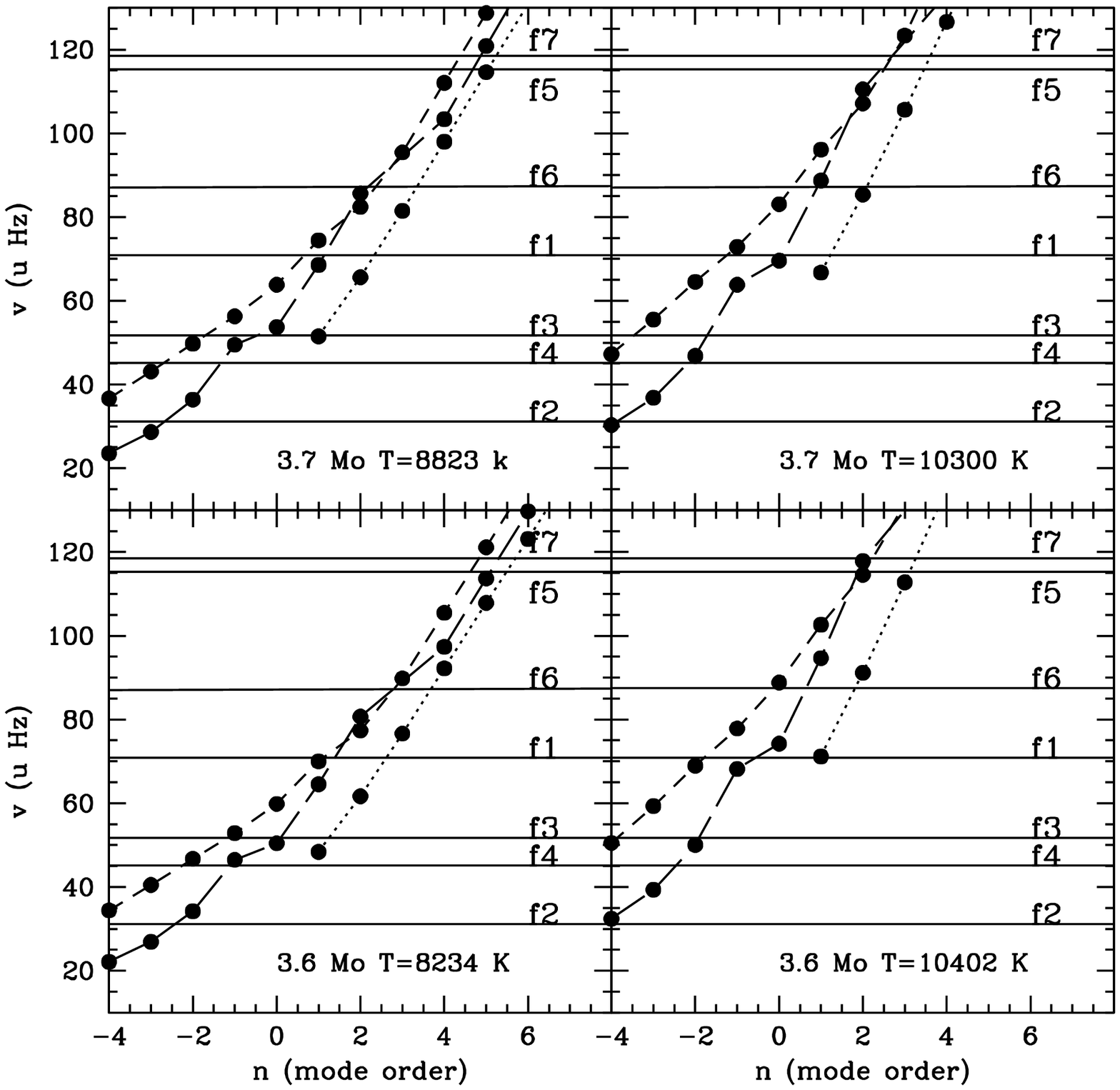}
      \caption{The same as for Fig.~\ref{fig7} but for 3.8~$M_{\odot}$
         (bottom panels)
        and 3.9~$M_{\odot}$ (upper panels).}
      \label{fig9}
   \end{figure}

\begin{figure}
   \centering
   \includegraphics[width=9cm]{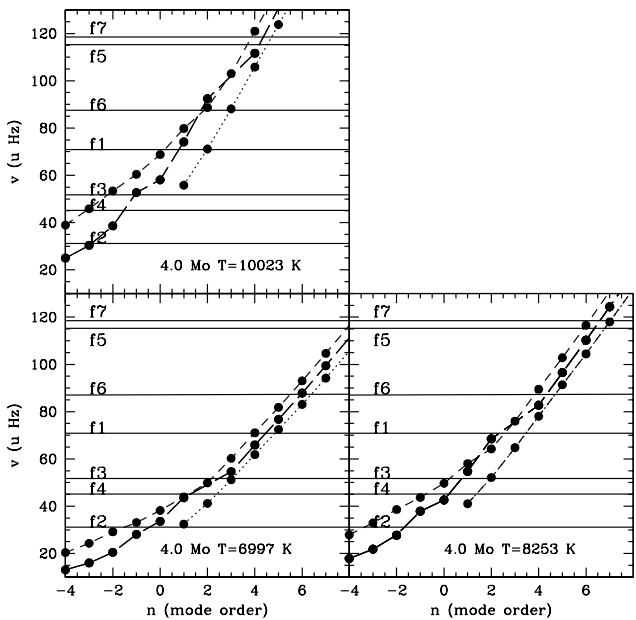}
      \caption{The same as for Fig.~\ref{fig7} but for 4~$M_{\odot}$.}
      \label{fig10}
   \end{figure}

 For each model, frequencies for $l$=0,1,2 \footnote{The reason
 why we consider only l=0,1,2 modes is that, as noticed by other
 authors (see e.g. Suran et al. 2001, Baglin et al. 2000), these modes
 are expected to be easily detectable, as their visibility
 coefficients remain large after integration over the whole stellar
 disk.} modes were computed using the Aarhus adiabatic nonradial
 pulsation code (ASTEC-{\tt http://astro.phys.au.dk/$\sim$jcd/adipack.n/})
 and compared with the observed periodicities. The range covered
 by the predicted frequencies depends on both the stellar mass and the
 effective temperature, moving toward higher values as the stellar
 mass (effective temperature) decreases (increases) at fixed
 effective temperatures (mass).  In Figs.~\ref{fig7}-\ref{fig10} the
 predicted frequencies with $l$=0 (filled symbols connected by dotted
 lines), $l$=1 (filled symbols connected by long dashed lines) and
 $l$=2 (filled symbols connected by dashed lines) are compared with
 the empirical periodicities (labelled horizontal solid lines) for the
 labelled masses and effective temperatures.

We find that the three lowest frequencies, namely $f_2$, $f_3$ and
 $f_4$, cannot be reproduced simultaneously by $p$ modes in the
 explored mass and effective temperature range.  In particular, $f_2$
 is only consistent with a $g$~mode ($l$=1,2 with $n<0$) unless
 the effective temperature is much lower than the estimated empirical
 range. To illustrate this point, we show in the left bottom panel of
 Fig.~\ref{fig10} the plot for a significantly cooler model
 ($T_\mathrm{eff}$=6997~K) with $M$=4.0~$M_{\odot}$. In this
 case, the range of the three lowest observed frequencies is covered
 with the predicted $l=0$, $n=0,1,2$ modes, with $f_2$ close to the
 theoretical result for the fundamental mode.  Pulsation in these low
 order radial modes at such a low effective temperature is also
 expected on the basis of the location of the predicted instability
 strip (see Fig.~\ref{fig6}).

 A tentative mode identification based on the various plots is
 summarized in Table~\ref{tab5}, available in electronic form. Model
 frequencies are required to match the observed ones within 2.5$\mu
 Hz$, in order to take into account both the mean error on the
 measured periodicities ($\sim$ 0.5 $\mu Hz$) and an estimate of the
 (unknown) model intrinsic uncertainty.  The models reproducing all
 the observed frequencies with $l$=0,1,2 $p$ and/or $g$ (or $f$) modes
 are the ones reported in Table \ref{bestfit} and shown in
 Fig.~\ref{fig6} as filled circles. The stellar mass and
 luminosity associated with these models range from 3.6 to 4.0
 $M_{\odot}$ and from 125.2 to 207.6 $L_{\odot}$. To
 remove the degeneracy on the model solution, an independent empirical
 estimate of the large separation would be very important. However, the
 evaluation of the large separation from the frequency power spectrum
 would require more accurate, long-time data.

\begin{center}
\begin{table}[h]
\renewcommand{\arraystretch}{1.1}
\caption{Preliminary mode identification as a function of the explored
model input parameters for models reproducing all the seven observed
frequencies. The mass is in solar units ($M_\odot$) and effective
temperature is in K.
\label{bestfit}}
\begin{tabular}{lc|c|rrrrrrr}
\hline\hline
{\bf Model} &  &  &
      $f_1$ & $f_2$ & $f_3$ & $f_4$ & $f_5$ & $f_6$&$f_7$ \\
Mass & $T_\mathrm{eff}$ & $l$ & & \multicolumn{4}{c}{Possible value of $n$} \\
\hline\hline
{\bf mod24} &   & 0 & 1  &    &    &    &    & &\\
  3.6 & 10\,402 & 1 &    & -4 &  -2&    &    & &2\\
      &         & 2 & -2 & -8 &  -4& -5 & 2  & 0&  \\
\hline
{\bf mod25} &   & 0 & 3   &      &    & 1  &  6  & 4&6 \\
  3.7 & 8\,211  & 1 &     &  -2  &    &-1  &    & &\\
      &         & 2 & 2   & -4   & -1 & -2 & 5  & 3 &  \\
\hline
{\bf mod26} &   & 0 &    &    & 1    &    & 5  & &\\
  3.7 & 8\,823  & 1 & 1  & -3 & -1   &    &    & 2&5 \\
      &         & 2 &    & -5 &  -2  &  -3&    & &\\
\hline
{\bf mod29} &   & 0 &  3  &    &    &  1  &    & &\\
  3.8 & 8\,155  & 1 &  2  &  -2&    &    &  6  & 4&6 \\
      &         & 2 &  2  & -4 & 0   & -1   &    & &\\
\hline
{\bf mod33} &   & 0 &    &    & 2   &    &    & &\\
  3.9 & 8\,210  & 1 & 2  & -2 &    &    & 6   & 4&   \\
      &         & 2 &    & -4 & 0  & -1 &    & &6\\
\hline
{\bf mod37} &   & 0 & 5  &  1 & 3  &    & 9  & &9\\
  4.0 & 6\,997  & 1 &    &    &    & 1  &    & 6& \\
      &         & 2 & 4  &  -2& 2  & 1  & 8  & &8\\
\hline
{\bf mod38} &   & 0 &    &    & 2  &    &    & &7\\
  4.0 & 8\,253  & 1 & 2  &    &    &    &    & &\\
      &         & 2 &    & -3 & 0   & -1 & 6  & 4&6  \\
\hline\hline
\end{tabular}
\end{table}
\end{center}

We note that, among the best-fit  models,  mod37 is inside the 
theoretical instability
 strip for the three lowest radial modes, so that $f_2$ and $f_3$ are
 correctly reproduced by pulsation in the fundamental and the second
 overtone, respectively. 
In this case
 all the seven observed frequencies can indeed be reproduced by $p$ modes.
 Therefore, the results seem to suggest that either the effective
 temperature found in the literature is significantly
 overestimated, or that we have to admit the possibility of pulsation in at 
 least one $g$ mode.

Available studies point towards the existence of unstable low
degree p-modes in PMS stars.  Consequently, in the above
analysis, models with $p$-modes in the observed frequency
range have been preferred.  The possibility that $g$-modes are also
excited to the observed amplitudes has yet to be confirmed by a
detailed theoretical analysis of the excitation mechanisms of such
modes in young stars and supported by observational mode
identification.  A more detailed analysis (theoretical and
observational) in this direction is clearly required.
 
\subsection{Effect of rotation on mode identification}

Here, we give a simplified, preliminary account of the effect of
rotation on nonradial modes using the asymptotic relation for the
rotational displacement. This is given by $m f_{rot}$, where $m$ is
the spherical harmonic number and $f_{rot}=(v_{rot} \sin i) / (2 \pi R
\sin i)$ the rotational frequency.  The term $v_{rot} \sin i$ can be
taken from published values. According to Sect. 3, the rotational
velocity varies between a minimum of $v_{rot} \sin i \sim$85 km/s and
a maximum of 200 km/s.  For the $\sin i$ term in the denominator, we
need to make an assumption on the inclination angle $i$.  For
simplicity, we consider two extreme cases, namely $i_1=\pi/4$ and
$i_2=\pi/2$. Finally, for the theoretical radius we take the values of
our best-fit models as given in Table \ref{bestfit}.  Adopting
$v_{rot} \sin i=$200 km/s, we find a rotational splitting $8.5(17) \le
\Delta f_1 \le 14.6(29.2) \mu$Hz and $6(12) \le \Delta f_2 \le
10.3(20.6) \mu$Hz for $m=\pm 1(2)$, respectively.  Conversely, for a
velocity of 85 km/s the splittings are reduced to $3.6(7.2) \le \Delta
f_1 \le6.2(12.4) \mu$Hz and $2.5(5.0) \le \Delta f_2 \le 4.4(8.8)
\mu$Hz for $m=\pm1(2)$, respectively.  We have also calculated the
upper limit of $\Delta f$ under the extreme assumption of $v <
v_{breakup}$, where $v_{breakup}$ is the break up velocity for the
models reported in Table \ref{bestfit}, yielding $\Delta f_max \sim 20
\mu$Hz and thus suggesting that the rotational splitting might
introduce significant error bars on the theoretical mode
identification.

On the other hand, if we assume that the measured periodicity at 2.2 d is
associated with rotation, we derive a rotational splitting of $\simeq
\pm 5(10) \mu Hz$ for $m=\pm 1(2)$, in better agreement with the
values $m \times 2.5$ to $m \times 4.4 \mu Hz$ obtained
for a rotational velocity of 85 km/sec and a radius range covered by
the best fit models reported in Table \ref{bestfit} 
and a star seen close to edge-on ($i=\pi/2$).  Thus, these initial
results seem to indicate that rotation may affect significantly  the
theoretical mode identification, with larger radii providing smaller
rotational splitting for a given rotational velocity.  However, since rotation
alters the structural and evolutionary properties of stellar models, an
accurate estimate of the effects on mode identification can only be obtained 
by a specifically designed numerical code that incorporates all the relevant
physical ingredients (see e.g. Reese et al. 2006).

\section{Conclusions}

We have detected $\delta$ Scuti-type pulsation in the Herbig Ae star
VV Ser on the basis of ground-based observations carried out during
three consecutive years. In particular, using the best dataset obtained in 2004
seven pulsation frequencies have been measured  with
values ranging from 2.69 to 10.24 $c/d$. 
These results allow to include
VV Ser in the list of known PMS stars with $\delta$ Scuti 
variations ( \cite{zwintz}, \cite{ripepi05}). In order to compare the
extracted periodicities with model predictions, we have considered a
large empirical range both in luminosity and effective temperature,
reflecting the existing uncertainties on the stellar properties. 
Within this space of physical
parameters, we have computed a fine grid of inner structure models
along the CESAM evolutionary tracks. The corresponding pulsation
frequencies for $l$=0,1,2 modes have been evaluated by means of the
ASTEC code and compared with the
observed ones by requiring an agreement within 2.5 $\mu Hz$.

As a result, we find a number of best fit models corresponding to PMS
stars with mass 3.6--4.0 $M_{\odot}$ and luminosity
$\log{L/L_{\odot}}\approx$2.1--2.3 that reproduce the seven observed
periodicities with radial and/or nonradial $p$ and $g$ modes.
  Such an independent confirmation of the fundamental stellar
  parameters range previously estimated for VV Ser is an important
  byproduct of the main objective of asteroseismology (which will be
  addressed in a future work, requiring more and better observations
  and more accurate modelling), i.e. of sensing the internal structure
  of the star.  We also notice that if the effective temperature is
significantly lower ($\sim$ 7000 K) than the values based on the
empirical spectral types ($\simeq$ 9000 K), we are able to reproduce
all the observed periodicities with $p$ modes for a stellar mass of 4
$M_{\odot}$ and a luminosity of 125 $L_{\odot}$. An accurate spectral
type determination of VV Ser is urgently needed in order to
discriminate among the different solutions. In the meantime, the
present analysis seems to indicate a stellar mass in the range of the
largest masses expected (and indeed observed) for this class of
pulsators.  The possible effect of rotation on the predicted
frequencies depends on the adopted estimates of the rotational
velocity and inclination angle, as well as on the stellar radius. If
the measured periodicity at 2.2 d is interpreted in terms of rotation,
the corresponding rotational splitting is of the order of $m \times 5
\mu Hz$.  Finally, we note that further photometric observations of VV
Ser, especially by means of multisite campaign, would be of great help
to better define the frequency spectrum and to eliminate the
uncertainty on the observed frequencies due to the 1 c/d alias.


\begin{acknowledgements}
We thank the referee, T. B\"ohm, for useful comments that
greatly improved the presentation.  V.R. wish to thank the personnel of
Loiano Observatory for their competent and kind help during the
observations. Partial financial support for this work was provided by
PRIN-INAF 2005 under the project ``Stellar Clusters as probes of
stellar formation and evolution'' (P.I. F. Palla). This research has
made use of the SIMBAD database, operated at CDS, Strasbourg, France.
MJM and JPM were supported in part by FCT through project {\scriptsize
POCTI/CTE-AST/57610/2004} from POCTI, with funds from the European
programme FEDER.

\end{acknowledgements}

\bibliographystyle{}

\Online
\appendix

\begin{table}
\caption{Log of the observations.}
\label{tab1}
\begin{tabular}{ccc}
\hline
\noalign{\smallskip}
HJD-2450000 & HJD-2450000  & Duration  \\
  start (days) &  end (days)&  (hours)    \\
\noalign{\smallskip}
\hline
\noalign{\smallskip}
52488.362&	52488.517&       3.7 \\
52490.338&	52490.510&       4.1 \\
52492.344&	52492.506&       3.9 \\
52493.335&	52493.490&       3.7 \\
52800.427&	52800.583&       3.7 \\
52801.410&	52801.574&       3.9 \\
52802.425&	52802.578&       3.7 \\
52803.423&	52803.573&       3.6 \\
53182.414&	53182.514&       2.4 \\
53183.371&	53183.482&       2.7 \\
53184.356&	53184.609&       6.1 \\
53185.350&	53185.597&       5.9 \\
53186.420&	53186.600&       4.3 \\
53209.402&	53209.544&       3.4 \\
53210.399&	53210.518&       2.9 \\
\noalign{\smallskip}
\hline
\end{tabular}
\end{table}

\begin{table}[h]
\caption{Physical properties of the selected PMS models. \label{tab4}}
\begin{tabular}{lcccrr}
\hline\hline\noalign{\smallskip}
Model & Mass & Radius & Luminosity & $T_\mathrm{eff}$ & Age \\
      & $\rm M/M_{\odot}$ & $\rm R/R_{\odot}$ & $\rm L/L_{\odot}$ &K & Gy \\
\hline\noalign{\smallskip}
mod1  &2.2 &1.99 &20.74  &8749  &5.82 \\
mod2  &2.2 &1.77 &27.27  &9913  &6.98 \\
mod3  &2.4 &2.09 &29.46  &9304  &4.65 \\
mod4  &2.4 &2.04 &37.54  &10000 &5.11 \\
mod5  &2.6 &3.54 &57.56  &8459  &3.24 \\
mod6  &2.6 &3.14 &66.90  &9319  &3.38 \\
mod7  &2.6 &2.18 &43.08  &10023 &4.09 \\
mod8  &2.8 &4.08 &66.50  &8168  &2.63 \\
mod9  &2.8 &3.23 &90.99  &9885  &2.85 \\
mod10  &3.0 &4.58 &77.3  &8007  &2.17 \\
mod11  &3.0 &4.17 &91.9  &8758  &2.17 \\
mod12  &3.0 &3.74 &107.9 &9621  &2.35 \\
mod13  &3.0 &3.32 &122.2 &10546 &2.45 \\
mod14  &3.2 &4.67 &105.7 &8573  &1.88 \\
mod15  &3.2 &4.22 &125.5 &9417  &1.96 \\
mod16  &3.2 &3.75 &146.2 &10375 &2.04 \\
mod17  &3.4 &5.37 &110.1 &8086  &1.55 \\
mod18  &3.4 &4.96 &130.4 &8762  &1.61 \\
mod19 &3.4 &4.53 &152.5 &9539  &1.67 \\
mod20 &3.4 &4.08 &176.0 &10414 &1.73 \\
mod21 &3.6 &5.71 &134.7 &8234  &1.33 \\
mod22 &3.6 &5.31 &156.4 &8869  &1.38 \\
mod23 &3.6 &4.88 &180.9 &9590  &1.43 \\
mod24 &3.6 &4.44 &207.6 &10402 &1.48 \\
mod25 &3.7 &5.95 &144.3 &8211  &1.23 \\
mod26 &3.7 &5.41 &167.0 &8823  &1.28 \\
mod27 &3.7 &5.12 &192.9 &9518  &1.32 \\
mod28 &3.7 &4.68 &221.2 &10300 &1.36 \\
mod29 &3.8 &6.21 &152.9 &8155  &1.14 \\
mod30 &3.8 &5.81 &175.9 &8727  &1.18 \\
mod31 &3.8 &5.40 &202.2 &9376  &1.22 \\
mod32 &3.8 &4.97 &231.2 &10106 &1.26 \\
mod33 &3.9 &6.39 &166.5 &8210  &1.06 \\
mod34 &3.9 &6.00 &190.7 &8766  &1.09 \\
mod35 &3.9 &5.59 &218.1 &9392  &1.13 \\
mod36 &3.9 &5.16 &248.4 &10092 &1.16 \\
mod37 &4.0 &7.63 &125.2 &6997  &0.89 \\
mod38 &4.0 &6.58 &180.5 &8253  &0.98 \\
mod39 &4.0 &6.20 &205.2 &8760  &1.02 \\
mod40 &4.0 &5.80 &232.9 &9366  &1.05 \\
mod41 &4.0 &5.40 &263.8 &10023 &1.08 \\
\noalign{\smallskip}
\hline \hline
\end{tabular}
\end{table}

\begin{longtable}{lc|c|rrrrrrr}
\caption{Preliminary mode identification as a function of the explored
model input parameters. The mass is in solar units ($M_\odot$) and 
the effective temperature in K.
\label{tab5}}\\
\hline\hline
{\bf Model} &  &  &
      $f_1$ & $f_2$ & $f_3$ & $f_4$ & $f_5$ & $f_6$&$f_7$ \\
Mass & $T_\mathrm{eff}$ & $l$ & & \multicolumn{4}{c}{Possible value of $n$} \\
\hline
\endfirsthead
\caption{continued.}\\
\hline\hline
{\bf Model} &  &  &
      $f_1$ & $f_2$ & $f_3$ & $f_4$ & $f_5$ & $f_6$&$f_7$ \\
Mass & $T_\mathrm{eff}$ & $l$ & & \multicolumn{4}{c}{Possible value of $n$} \\
\hline
\endhead
\hline
\endfoot
{\bf mod1} &    & 0 &    &      &    &    &   & &\\
  2.2 & 8\,749  & 1 &    &  -5  & -3 &    &   & &\\
      &         & 2 &    &      &    &    &   & &\\
\hline
{\bf mod2} &    & 0 &    &      &    &    &   & &\\
  2.2 & 9\,913  & 1 &    &      & -4 & -5 &   &-2& \\
      &         & 2 & -5 &  -7  & -7 & -9 &   & -4&\\
\hline
{\bf mod3} &    & 0 &    &      &    &    &   & &\\
  2.4 & 9\,304  & 1 &    & -5   &    & -3 &   & &\\
      &         & 2 &    & -9   & -5 & -6 &   & &\\
\hline
{\bf mod4} &    & 0 &    &      &    &    &   & &\\
  2.4 & 10\,000 & 1 &    & -6,-7&    & -4 &   & &\\
      &         & 2 &    &      &    & -8 &   & &-2 \\
\hline
{\bf mod5} &    & 0 &    &      &    &    &   & 1&\\
  2.6 & 8\,459  & 1 &    &      &    &    & 1 & -1& \\
      &         & 2 &    &      &    &    &   & -2&\\
\hline
{\bf mod6} &    & 0 &    &      &    &    &   &  &\\
  2.6 & 9\,319 & 1  &    &      &    &    &   & &\\
      &         & 2 &    &      &    &    &-1 & &\\
\hline
{\bf mod7} &    & 0 &    &      &    &    &   & & \\
  2.6 & 10\,023 & 1 &    &      &    &    &   & &\\
      &         &   &    &      &    &    &   & &\\ 
\hline
{\bf mod8} &    & 0 & 1   &      &    &    &3   & & \\
  2.8 & 8\,168  & 1 &-1   &      &    &    &   & &2\\
      &         &   & -2  &      &    &    &   & &2\\ 
\hline
{\bf mod9} &    & 0 &    &      &    &    &   & &\\
  2.8 & 9\,885 & 1 &     &       &    &   &   & &\\
      &         &   &    &      &    &    &   & &\\ 
\hline
{\bf mod10} &    & 0 &    &      &    &    &   & &4\\
  3.0 & 8\,007  & 1 &    &  -4  &    & -2   &  & & \\
      &         & 2 & -1 &  -7  &    & -4   & 3& &\\
\hline
{\bf mod11}  &   & 0 & 1  &      &    &    & 3   & &\\
  3.0 & 8\,758  & 1 & -1 &  -5  &  -2&    &    & &2\\
      &         & 2 & -2 & -8,-9&    &  -5 &    &&2 \\
\hline
{\bf mod12} &    & 0 &    &      &    &    &    & & \\
  3.0 & 9\,621  & 1 &    &   -6 &    &    &    & &\\
      &         & 2 & -3 &  -10 &  -5&  -6&    & &\\
\hline
{\bf mod13} &    & 0 &    &       &    &    &    & & \\
  3.0 & 10\,547 & 1 &    & -6    & -3 & -4 &    & &\\
      &         & 2 & -4 & -11   & -6 & -7 &    & &\\
\hline
{\bf mod14} &    & 0 &    &    &    &    &    & &4\\
  3.2 & 8\,573  & 1 &    & -4 &    & -2 &    & &\\
      &         & 2 & -1 & -7 & -3 & -4 & 3  & &\\
\hline
{\bf mod15} &    & 0 & 1   & -5   &    &    &  3 && \\
  3.2 & 9\,417  & 1 & -1  & -9   & -2 &    &    & &2\\
      &         & 2 &  2  &  -8  &  -4&  -5&    & &2\\
\hline
{\bf mod16} &    & 0 &    &    &    &    &    & 1 &\\
  3.2 & 10\,375 & 1 &    & -5 &    &    &  1 & &\\
      &         & 2 & -3 & -10&-5  &  -6&    && \\
\hline
{\bf mod17}  &   & 0 &    &    & 1    &    & 5  & &\\
  3.4 & 8\,086  & 1 &  1 & -3 & -1   &    &    & 2 & \\
      &         & 2 &    &  -5&  -2  &  -3&    & &\\
\hline
{\bf mod18}  &   & 0 &    &    &    &    &    & &\\
  3.4 & 8\,762  & 1 &    &  -3&    &    &  4  & &4\\
      &         & 2 &    &-6,-7& -3& -4 &    & &\\
\hline
{\bf mod19} &   & 0 &    &    &    &    &    &  2& \\
  3.4 & 9\,538  & 1 &    & -4 &    &    &    & 1 &\\
     &          & 2 &    & -7,-8   &    &    &    && \\
\hline
{\bf mod20} &   & 0 &    &    &    &    &    & &\\
  3.4 & 10\,414 & 1 &    & -5 &    &  -3&    & &\\
     &          & 2 &    & -9 &    &    & 1  &-1&  \\
\hline
{\bf mod21} &   & 0 &    &    &    &    &    & &\\
  3.6 & 8\,234  & 1 &    &    &    &  -1 &   5& &\\
     &          & 2 & 1  & -5 & -1  & -2 &    &  3& \\
\hline
{\bf mod22} &   & 0 & 2   &    &    &    &    & 3 &5 \\
  3.6 & 8\,868  & 1 & 1   &-3  & -1 &    &    & &\\
     &          & 2 &     &  -6&  -2& -3 &    & 2&4\\
\hline
{\bf mod23} &   & 0 &    &    &    &    &  4  & &4 \\
  3.6 & 9\,590  & 1 &    &    &    & -2 &    & &\\
      &         & 2 &    &  -7& -3 & -4 & 3  & 1&  \\
\hline
{\bf mod24} &   & 0 & 1  &    &    &    &    & &\\
  3.6 & 10\,402 & 1 &    & -4 &  -2&    &    & &2\\
      &         & 2 & -2 & -8 &  -4& -5 & 2  & 0&  \\
\hline
{\bf mod25} &   & 0 & 3   &      &    & 1  &  6  & 4&6 \\
  3.7 & 8\,211  & 1 &     &  -2  &    &-1  &    & &\\
      &         & 2 & 2   & -4   & -1 & -2 & 5  & 3 &  \\
\hline
{\bf mod26} &   & 0 &    &    & 1    &    & 5  & &\\
  3.7 & 8\,823  & 1 & 1  & -3 & -1   &    &    & 2&5 \\
      &         & 2 &    & -5 &  -2  &  -3&    & &\\
\hline
{\bf mod27} &   & 0 &    &    &    &    &    & &\\
  3.7 & 9\,518  & 1 &    & -3   &    &    &  4  && \\
      &         & 2 & 0   & -6  &    &    &   & &\\
\hline
{\bf mod28} &   & 0 &    &    &    &      &    &  2 &\\
  3.7 & 10\,300 & 1 &    & -4 &    &  -2  &    &  1 &\\
      &         & 2 & -1 & -7,-8 & & -4     &    & &\\
\hline
{\bf mod29} &   & 0 &  3  &    &    &  1  &    & &\\
  3.8 & 8\,155  & 1 &  2  &  -2&    &    &  6  & 4&6 \\
      &         & 2 &  2  & -4 & 0   & -1   &    & &\\
\hline
{\bf mod30} &   & 0 &    &    &    &    &    & &\\
  3.8 & 8\,727  & 1 &    &    &    & -1   &  5  & &\\
      &         & 2 & 1  & -5   & -1 & -2   &    &  3 &\\
\hline
{\bf mod31} &   & 0 & 2   & -3   &    &    &    &  3&5 \\
  3.8 & 9\,376  & 1 & 1   & -6   & -1 &    &    & &\\
      &         & 2 &    &  -5  & -2 & -3 &    & 2&4 \\
\hline
{\bf mod32} &   & 0 &    &    &    &    & 4  & &4\\
  3.8 & 10\,106 & 1 &    &    &    & -2 &    & &\\
      &         & 2 &    & -7 & -3 & -4 & 3   & 1&  \\
\hline
{\bf mod33} &   & 0 &    &    & 2   &    &    & &\\
  3.9 & 8\,210  & 1 & 2  & -2 &    &    & 6   & 4&   \\
      &         & 2 &    & -4 & 0  & -1 &    & &6\\
\hline
{\bf mod34} &   & 0 &    &    &    & 0  &    & 4&6  \\
  3.9 & 8\,766  & 1 &    & -2 &    & -1 &    & &\\
      &         & 2 &    & -4 & -1 & -2 & 5  & 3&5  \\
\hline
{\bf mod35} &   & 0 &    &    & 1  &    & 5  & &\\
  3.9 & 9\,392  & 1 & 1  & -3 & -1 &    &    &2&  \\
      &         & 2 &    &-5  & -2 & -3 & 4  & &\\
\hline
{\bf mod36} &   & 0 &    &    &    &    &    &  &\\
  3.9 & 10\,092 & 1 &    & -3 &    &    &    & &4\\
      &         & 2 & 0   & -6 &    &    &    && \\
\hline
{\bf mod37} &   & 0 & 5  &  1 & 3  &    & 9  & &9\\
  4.0 & 6\,997  & 1 &    &    &    & 1  &    & 6& \\
      &         & 2 & 4  &  -2& 2  & 1  & 8  & &8\\
\hline
{\bf mod38} &   & 0 &    &    & 2  &    &    & &7\\
  4.0 & 8\,253  & 1 & 2  &    &    &    &    & &\\
      &         & 2 &    & -3 & 0   & -1 & 6  & 4&6  \\
\hline
{\bf mod39} &   & 0 & 3  &    &    & 1  &   6&4 &  \\
  4.0 & 8\,760  & 1 &    & -2 &    &    &    & &6\\
      &         & 2 & 2  & -4 &    & -1 &    & &\\
\hline
{\bf mod40} &   & 0 &    &    & 1  &    &    & & \\
  4.0 & 9\,366  & 1 &    &    &    & -1 & 5   & &5\\
      &         & 2 & 1  & -5 &-1  & -2 &    & &\\
\hline
{\bf mod41} &   & 0 & 2  &    &    &    &    & 3& \\
  4.0 & 10\,023 & 1 &    & -3 & -1 &    &    & &\\
      &         & 2 & 0  &-6  & -2 & -3 &    & 2&4  \\
\hline\hline
\end{longtable}

\end{document}